\documentclass[preprint,secnumarabic,amssymb,letterpaper,endfloats]{revtex4}
\usepackage{amsmath}%
\usepackage{bm}%
\usepackage{hyperref}%
\usepackage{graphicx}

\begin{document}

\title{Epitaxial Thin Films of the Giant-Dielectric-Constant Material
CaCu$_3$Ti$_4$O$_{12}$ Grown by Pulsed-laser Deposition}

\author{W. Si, E. M. Cruz, and P. D. Johnson}

\address{Physics Department, Brookhaven National Laboratory,
Upton, NY 11973}

\author{P. W. Barnes and P. Woodward}

\address{Department of Chemistry, The Ohio State Uiversity, Columbus, Ohio 43210}

\author{A. P. Ramirez}

\address{Condensed Matter \& Thermal Physics Group, Los Alamos National
Laboratory, Los Alamos, NM 87545}

\date{\today}

\begin{abstract}
Pulsed-laser deposition has been used to grow epitaxial thin
films of the giant-dielectric-constant material
CaCu$_3$Ti$_4$O$_{12}$ on LaAlO$_3$ and SrTiO$_3$ substrates with
or without various conducting buffer layers. The latter include
YBa$_2$Cu$_3$O$_7$, La$_{1.85}$Sr$_{0.15}$CuO$_{4+\delta}$ and
LaNiO$_3$. Above 100K - 150K the thin films have a temperature
independent dielectric constant as do single crystals. The value
of the dielectric constant is of the order of 1500 over a wide
temperature region, potentially making it a good candidate for
many applications. The frequency dependence of its dielectric
properties below 100K - 150K indicates an activated relaxation
process.
\end{abstract}

\pacs{}

\maketitle

Ferroelectric or relaxor materials, such as (Ba, Sr)TiO$_3$ (BST)
\cite{Kotecki99}, SrTiO$_3$ (STO) \cite{HCLi98} and
KTa$_{1-x}$Nb$_x$O$_3$ (KTN) \cite{Samara00}, have been widely
studied for various electronics applications, including dynamic
random access memory(DRAM). This is because, in general, a high
static dielectric constant is only observed in either a
ferroelectric or a relaxor material. However, both types of
material show a peak in the dielectric constant as a function of
temperature, making it undesirable for many applications because
a strong temperature dependence may cause the related device to
fail with temperature variations in the environment. Recently
CaCu$_3$Ti$_4$O$_{12}$ (CCTO) was discovered to have an
extraordinarily high dielectric constant $\sim$ 80,000 at room
temperature with minimal temperature dependence above 100K.
\cite{Subramanian00,Ramirez00,Homes01} No ferroelectric lattice
distortion has been observed by either high-resolution X-ray
\cite{Ramirez00} or neutron powder diffraction
\cite{Subramanian00}, posing interesting questions : why does the
material have such a giant dielectric constant and why does the
latter decrease dramatically below 100K. It is therefore of
considerable interest both from a fundamental and an application
point of view to see whether one can make thin films of this
material and examine its properties. In this letter, we show that
high quality CCTO thin films can be epitaxially grown on various
substrates with or without conducting buffer layers. The low
frequency dielectric constant is lower than that of the single
crystal, but still $\sim$ 1500 over a wide temperature region,
making it a competitive candidate for many applications.
\begin{figure}[b]
\centering
\caption{$\theta - 2\theta$ scan of (a) a 6000\,{\AA} CCTO thin
film on LAO substrate and (b) a 6000\,{\AA} CCTO thin film on a
1500\,{\AA} LSCO electrode layer on STO substrate.} \label{fig1}
\end{figure}

The CCTO thin films were grown by pulsed laser deposition (PLD)
on LaAlO$_3$ (LAO) or STO substrates, as well as
YBa$_2$Cu$_3$O$_7$ (YBCO), La$_{1.85}$Sr$_{0.15}$CuO$_{4+\delta}$
(LSCO) and LaNiO$_3$ (LNO) buffer electrode layers. These
electrode layers were also deposited on the substrates $in situ$
by PLD before the deposition of the CCTO thin film. The CCTO
target was prepared by solid state reaction with the purity of the
powders confirmed by X-ray powder diffraction. A KrF excimer
laser (wavelength: 248\,nm) was used with an energy density of
$\sim$ 2.0\,J/cm$^2$ and a repetition rate of 5 Hz. The substrate
was heated to 720\,$^{\circ}$C (800\,$^{\circ}$C for the YBCO
layer). An oxygen pressure of 200\,mTorr was used during the
deposition for all materials. After deposition, the films were
cooled to room temperature at a rate of 60\,$^{\circ}$C per
minute. Au pads, 0.5\,mm$^2$ in area with a thickness of about
4000\,{\AA}, were thermally evaporated to form a parallel-plate
capacitor structure. Due to the difficulty of etching the CCTO
thin films while maintaining the film quality, contacts were made
to two of the top Au pads effectively, allowing the capacitance
to be measured in two capacitors connected in series by the
bottom electrodes. Dielectric constants and loss tangents were
measured by a Keithley 3330 LCZ meter from 200Hz to 100kHz.

Thin films grown directly on STO substrates do not have the
correct phase or may have more than one orientation of CCTO. This
is understandable since the lattice mismatch between STO and CCTO
is large. CCTO has a cubic structure with a lattice constant
$a$=7.36\,{\AA} ($a/2$=3.68\,{\AA}), while STO has a cubic
lattice constant $a$=3.905\,{\AA}. Thin films grown on the LAO
substrate have only one orientation. This is shown in Fig.1 (a) in
a $\theta - 2\theta$ X-ray diffraction scan. Only the ($l00$)
peaks from CCTO are present indicating that the out-of-plane
alignment is good. The substrate temperature and the oxygen
pressure were found to be crucial for single phase formation of
CCTO. Indeed thin films deposited at 780\,$^{\circ}$C or
20\,mTorr were found to include the (310) orientation. In order
to make dielectric measurements, various conducting perovskite
oxides were fabricated as the bottom electrodes. Shown in Fig.1
(b) is a $\theta - 2\theta$ scan of a 6000\,{\AA} CCTO thin film
on a 1500\,{\AA} LSCO buffer layer on the STO substrate. Only the
($00l$) peaks from both CCTO and LSCO layers are present in the
spectrum with peaks from the STO substrate, indicating both layers
grow with the $c$-axis perpendicular to the substrate surface. It
is worth pointing out that, with a LSCO buffer layer, the CCTO
thin films can also be grown epitaxially on a STO substrate.
\begin{figure}[t]
\centering
\caption{$\phi$ scan of the (220) peak from CCTO thin film and
(103) peak from LSCO buffer layer. This is a 6000\,{\AA}
CCTO/1500\,{\AA} LSCO bilayer thin film on STO substrate.}
\label{fig2}
\end{figure}

The in-plane epitaxy of the CCTO thin films were examined by
$\phi$ scans. Fig.2 shows the results of such a scan from the
same bilayer thin film as in Fig.1 (b). Sharp peaks appear with
90$^{\circ}$ intervals for both the CCTO and LSCO layers, and the
STO substrate, indicating the $a$ axis of the CCTO and LSCO
layers are well aligned with the substrate. Combined with the
$\theta - 2\theta$ scan, it shows we have an excellent epitaxy
between the CCTO thin film, the LSCO buffer layer and the STO
substrate. Thus we are in a position to study the properties of
CCTO in thin film format.
\begin{figure}[t]
\centering
\caption{Temperature dependence of (a) the dielectric constant
and (b) the loss tangent tan $\delta$ of a 6000\,{\AA} CCTO thin
film grown on LNO electrode layer on LAO substrate from 200Hz to
100kHz.} \label{fig3}
\end{figure}

Because LNO has a lattice constant closer to CCTO than YBCO and a
much better conductivity than LSCO, we made our measurement on
thin films of CCTO/LNO/LAO. The temperature dependence of the
dielectric constant and loss tangent for such a film at various
frequencies from 200Hz to 100kHz is shown in Fig.3 (a) and (b)
respectively. The dielectric constant has a fairly high value at
room temperature, but, unlike the single crystal data, with a
large frequency dependence. It decreases rapidly with decreasing
temperature. The loss tangent shows the same trend with
temperature. In fact both are very similar to the very first
report by Subramanian {\sl et al.} at high temperature.
\cite{Subramanian00} Below a temperature of 250K, the film
behaves almost exactly the same as a single crystal. The
dielectric constants at all frequencies have a constant value of
the order of 1500 until about 100K - 150K, when they decrease
rapidly to approximately 200. This value is different from that
of the single crystal, which is about 100. \cite{Homes01}
Obviously, the system undergoes a transition from a high
$\epsilon$ state at high temperature to a low $\epsilon$ state at
low temperature and the frequency dependence only takes place
around the transition temperature regime, which is about 100K -
150K. The transition takes place at higher temperatures at higher
frequencies. Correspondingly, the peak observed in the
temperature dependence of the loss tangent, shifts toward higher
temperature. Finally, it is interesting to note that the results
shown in Fig.3 are very similar to those from the bulk material
CdCu$_3$Ti$_4$O$_{12}$, which does not have the same giant
dielectric constant at room temperature as bulk CCTO.
\cite{Homes02}
\begin{figure}[t]
\centering
\caption{The log of the frequency versus inverse peak temperature
in loss measurement. Both the data from thin film and single
crystal are presented. A thermally activated behavior $f\sim
\exp(-U_0/kT)$ is clearly observed.} \label{fig4}
\end{figure}

The large dielectric constant of CCTO has been interpreted as an
extrinsic mechanism, which was assumed to come from the sample
microstructure such as boundary or interface effects. \cite{He02}
In fact, one recent paper has claimed that CCTO is a one-step
internal barrier layer capacitor. \cite{Sinclair02} With such a
model, one would conclude that the thin films have less defects
than the single crystals because of their lower dielectric
constant. This is a possiblility though it can often be the
reverse. Thin films, with their much reduced size in one
dimension, may have less planar defects, such as grain boundaries
and twin interfaces, than single crystals, particularly if the
latter have a large amount of intrinsic planar defects. Comparing
the data from the single crystals and the thin films, it seems
that the constant value of dielectric constant in a certain
temperature region and the decrease at low temperature is a
consistent property of the material, presenting an interesting
problem as to why the dramatic increase at even higher
temperature only shows up in thin films or polycrystalline
samples. \cite{Subramanian00} The dielectric constant also has a
significant dependence on the thickness of the CCTO thin films.
To date, thin films with a thickness less than 3000\,{\AA} have a
much lower dielectric constant ($\sim$700). However it should be
possible to improve this by further optimizing the growth
conditions, such as oxygen pressure and cooling rate after the
deposition or different conducting bottom electrode layer. These
studies as well as thickness dependence of the dielectric
properties will be published elsewhere.

Finally, it is worth pointing out that the loss tangent in the
thin film is actually lower than in the single crystal. This may
well be a reflection that the single crystals have more defects
than the thin films. Since the low temperature decrease of the
dielectric constant towards 100 or 200 is certainly associated
with the peak observed in the temperature dependence of the loss
tangent, this might just be due to the fact that the constant
value ($\sim$1500) of the dielectric constant in the thin film is
much lower than in the single crystal ($>$10,000). A precise
description should involve an understanding of the observed peak.
Homes {\sl et al.} used a Debye relaxation model to describe the
frequency dependence of the dielectric constant and find a
thermally activated behavior with an activation energy of about
630K in single crystal. \cite{Homes01} Here we show a simple way
to get a similar result from our loss tangent data. In Fig.4, we
plotted the frequency on a logarithmic scale versus the inverse
peak temperature in our loss measurement. Along with the thin
film data we also show the data from a single crystal. Clearly a
linear dependence is observed, indicating that $f \sim \exp
(-U_0/kT)$, in both sets of data. U$_0$ for the single crystal is
690K which is about the same as the value from fitting to the
Debye relaxation model for the dielectric constant.
\cite{Homes01} U$_0$ for the thin film is 880K. The difference
may well reflect the fact that the thin film is under strain due
to the lattice mismatch. Of course, the microstructure and
defects and domain difference would also play a role. Further
study on the evolution of the activated relaxation with thin film
thickness will be conducted.

In conclusion, we have successfully grown epitaxial CCTO thin
films by pulsed laser deposition. Thin films have similar
properties to the previously studied single crystals. The
dielectric constant is above 1500 within a fairly large
temperature region, making it a very attractive material for many
applications. Further investigations including the thickness
dependence and electric field dependence are under way.

Helpful discussions with Myron Strogin, Genda Gu, Chris Homes and
Tom Vogt are gratefully acknowledged. This work was supported in
part by the Department of Energy under Contract No.
DE-AC02-98CH10886.

\newpage \centering \printfigures
\newpage \centering \includegraphics[width=6.5in]{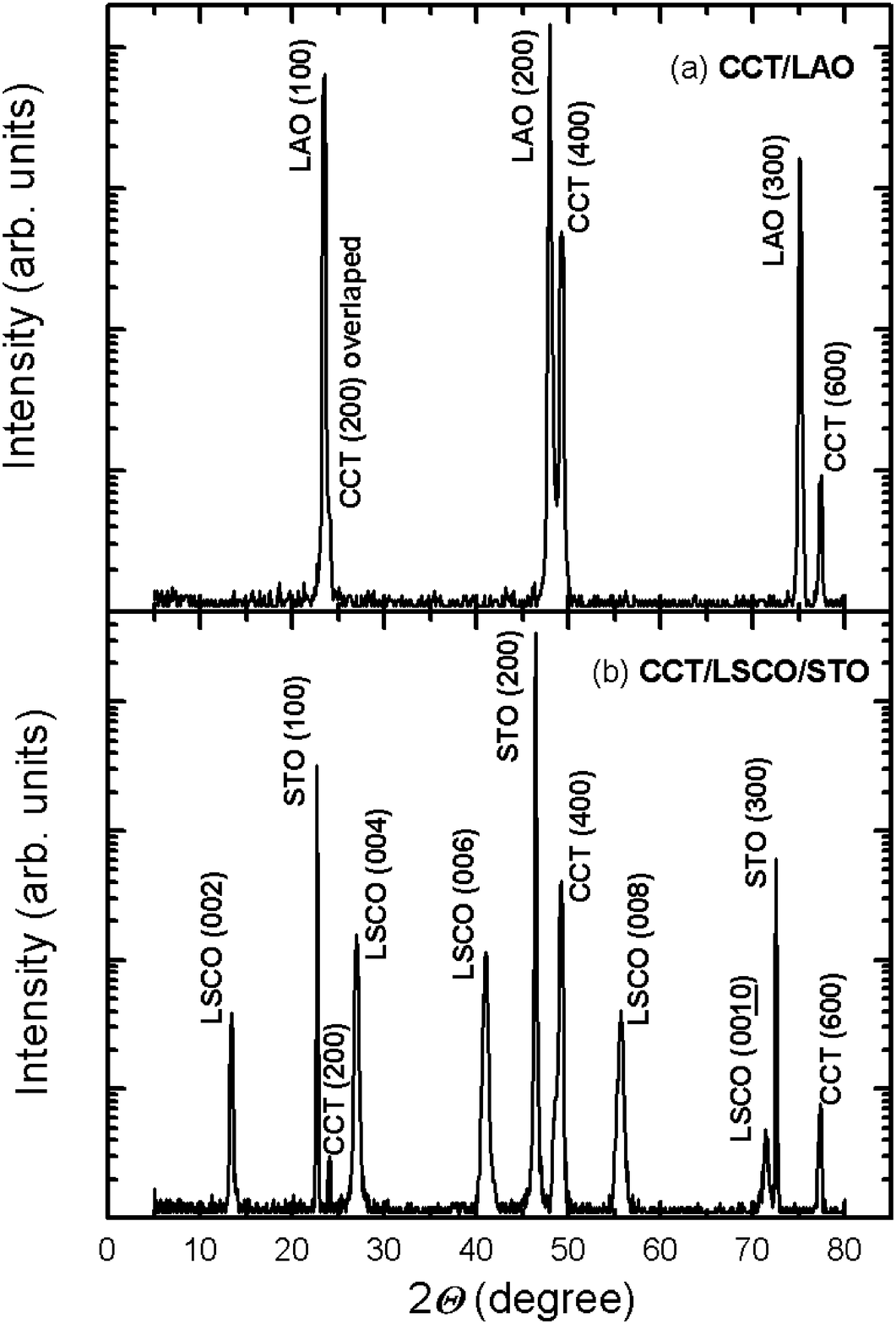}
\newpage \centering \includegraphics[width=6.5in]{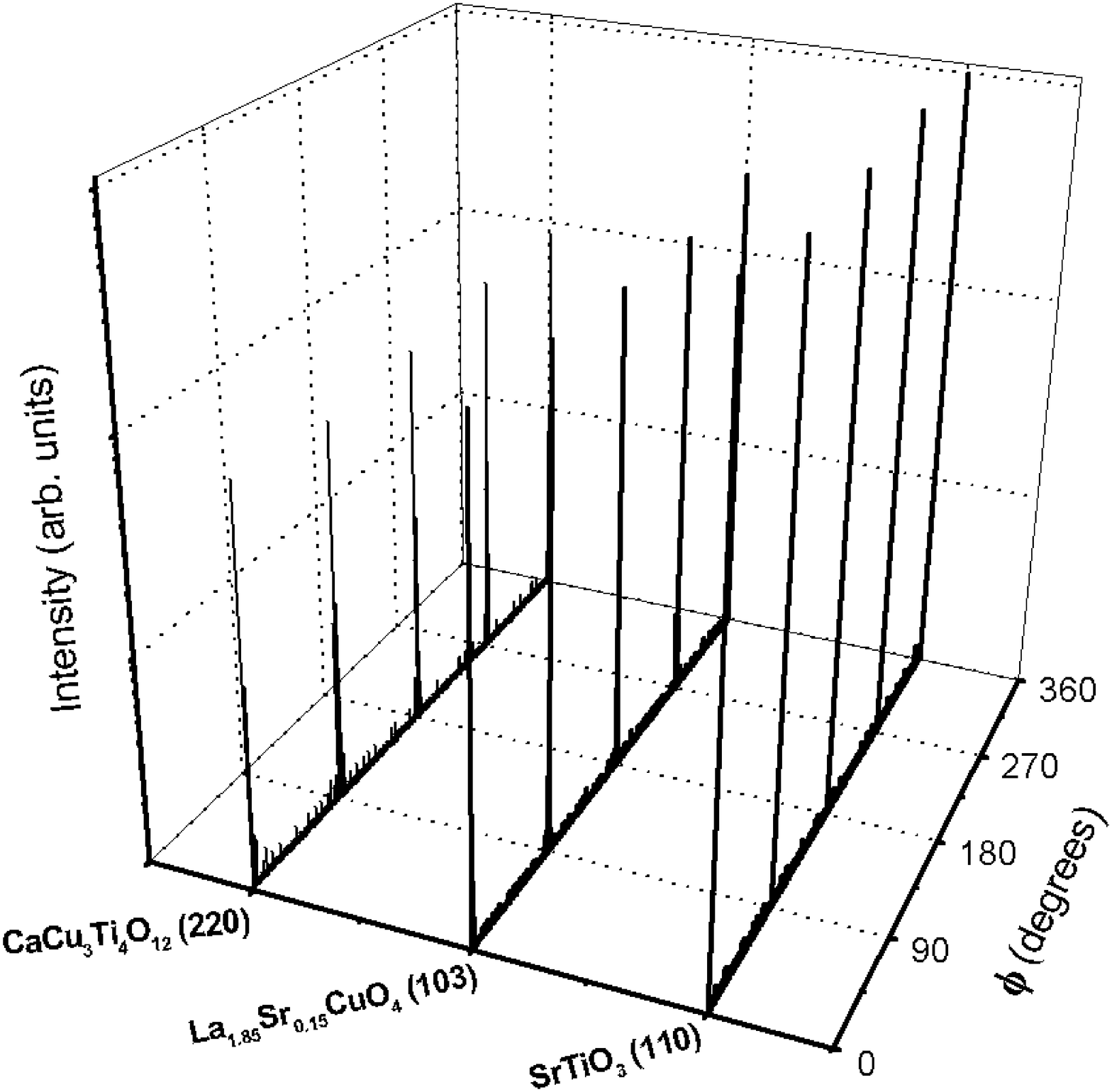}
\newpage \centering \includegraphics[width=6in]{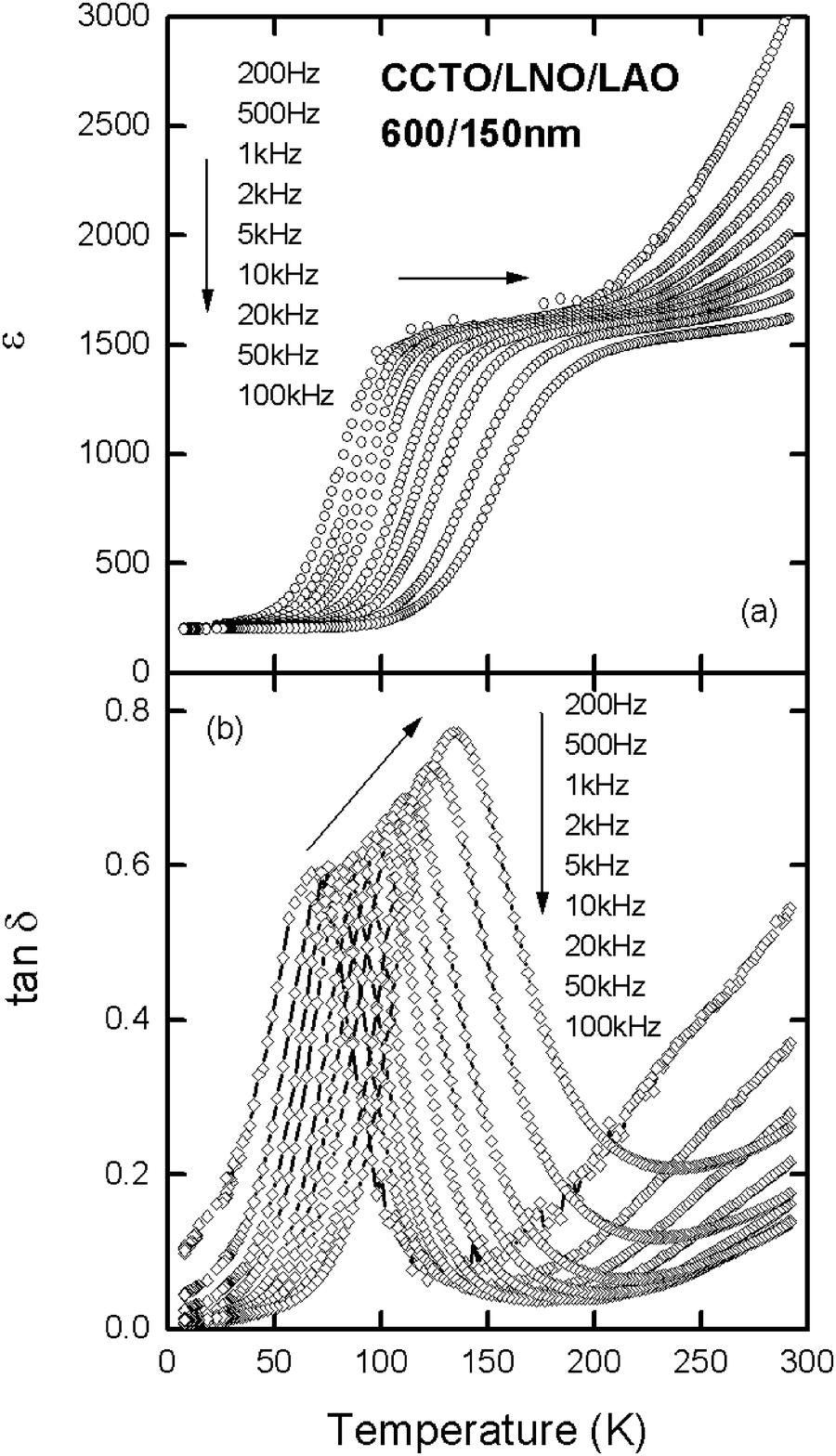}
\newpage \centering \includegraphics[width=6.5in]{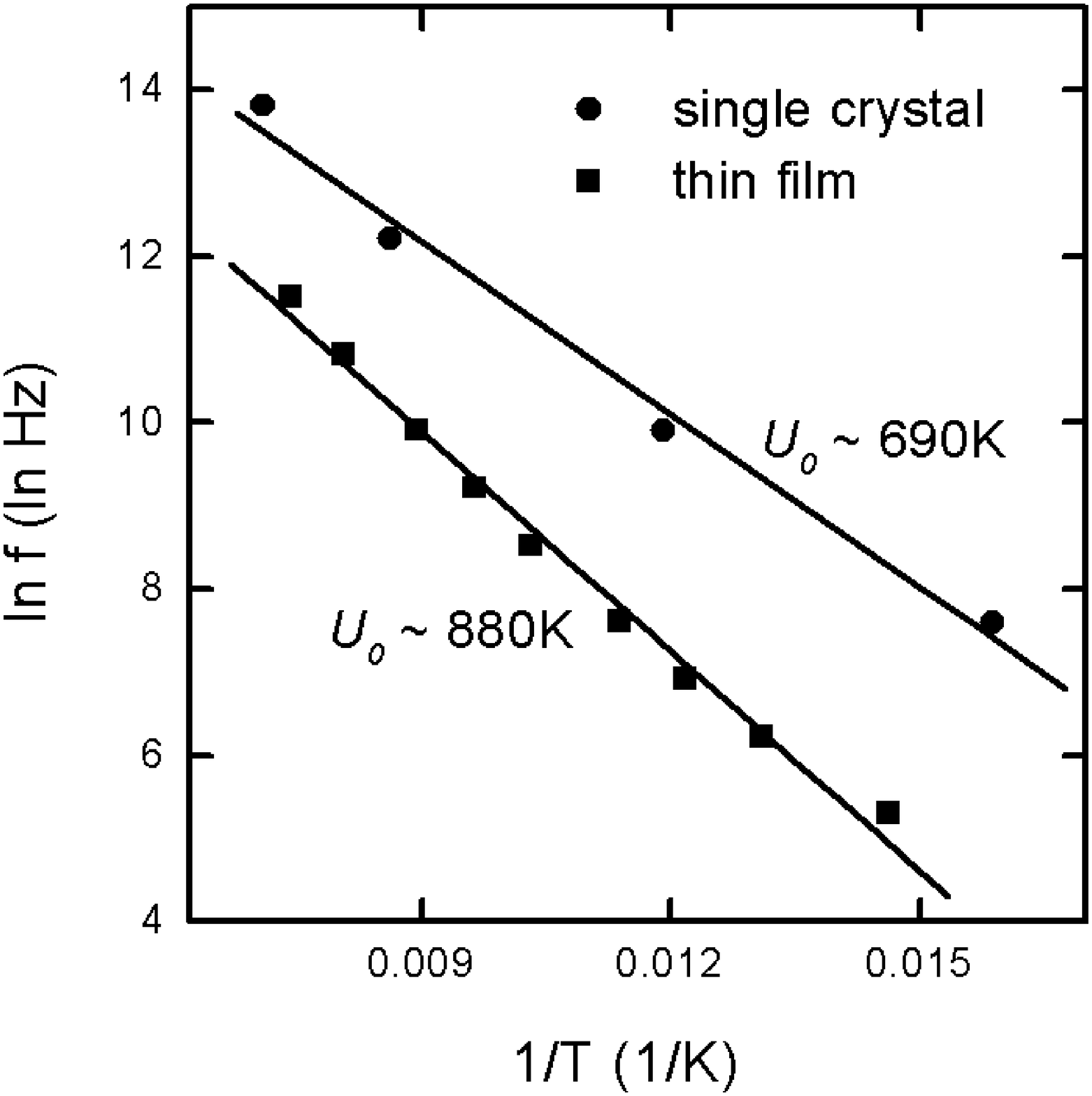}

\end{document}